\documentclass[12pt]{article}
\usepackage{graphicx}
\usepackage{xcolor}

\newcommand\pubnumber{}
\newcommand\pubdate{TTK-22-05, P3H-22-004 \\ \today}

\def\institute{Institute for Theoretical Particle Physics and Cosmology, RWTH-Aachen University, Aachen Germany}

\def\Title#1{\begin{center} {\Large #1 } \end{center}}
\def\Author#1{\begin{center}{ \sc #1} \end{center}}
\def\Address#1{\begin{center}{ \it #1} \end{center}}

\newcommand\pubblock{\rightline{\begin{tabular}{l} \pubnumber\\
         \pubdate  \end{tabular}}}
\newenvironment{Abstract}{\begin{quotation}  }{\end{quotation}}
\newenvironment{Presented}{\begin{quotation} \begin{center} 
             PRESENTED AT\end{center}\bigskip 
      \begin{center}\begin{large}}{\end{large}\end{center} \end{quotation}}

\def\beq{\begin{equation}}
\def\eeq#1{\label{#1}\end{equation}}
\def\eeqn{\end{equation}}

\def\beqa{\begin{eqnarray}}
\def\eeqa#1{\label{#1}\end{eqnarray}}
\def\eeqan{\end{eqnarray}}

\let\bar=\overbar

\def\Dslash{\not{\hbox{\kern-4pt $D$}}}
\def\dslash{\not{\hbox{\kern-2pt $\del$}}}

\def\msb{{\bar{\ssstyle M \kern -1pt S}}}

\usepackage[verbose,colorlinks=true,naturalnames=true,linkcolor=blue,urlcolor=blue, citecolor=blue]{hyperref}

\usepackage{floatrow}
\newfloatcommand{capbtabbox}{table}[][\FBwidth]

\begin{document}
\begin{titlepage}
\pubblock

\vfill
\Title{$t\overline{t}W^{\pm}$ at NLO in QCD:\\
Full off-shell effects and precision observables}
\vfill
\Author{ Jasmina Nasufi }
\Address{\institute}
\vfill
\begin{Abstract}
In light of recent tension between theory and experiment for the process $t\overline{t}W^{\pm}$ in the $3\ell$ channel, we present a phenomenological study of carefully chosen observables. We employ the full off-shell results at NLO QCD for $t\overline{t}W^{+}$ and $t\overline{t}W^-$ to build the cross section ratio, an observable which is expected to exhibit enhanced perturbative stability. Furthermore, we also revisit the charge asymmetries of the top and its decay products at the fiducial integrated level. The impact of \textit{modelling} is investigated for all observables. 
\end{Abstract}
\vfill
\begin{Presented}
$14^\mathrm{th}$ International Workshop on Top Quark Physics\\
(videoconference), 13--17 September, 2021
\end{Presented}
\vfill
\end{titlepage}
\def\thefootnote{\fnsymbol{footnote}}
\setcounter{footnote}{0}

\section{Introduction}

The associated production of a top-quark pair with a $W^{\pm}$-gauge boson can be used to constrain top-quark intrinsic properties, to probe physics beyond the Standard Model (SM) and it is an important background in searches with final states involving multiple leptons and b-jets at the LHC. Recent measurements and comparisons to theory data have been carried out for $t\overline{t}H$ and $t\overline{t}W^{\pm}$ in the three lepton final state by the ATLAS collaboration \cite{atlas1}. Comparing with theory results, they note a tension in the overall normalisation and the modelling of the process. This observation adds to previous ATLAS and CMS studies that report measuring an excess of events when comparing to current SM predictions \cite{atlas2}-\cite{atlas3}. \\
A concerted effort from the theory community in the past years, has provided a multitude of results with the aim to describe the process $pp\rightarrow t\underline{t}W^{\pm}$ and its decays more precisely. The full off-shell computation at NLO QCD was presented in \cite{paper1} and \cite{paper2}, followed by the complete NLO SM corrections \cite{paper3} shortly after. \\
In this proceeding we summarise results on the cross section ratio $\mathcal{R} = \sigma_{t\overline{t}W^{+}}^{NLO}/\sigma_{t\overline{t}W^{-}}^{NLO}$ and the charge asymmetries for the top-quark and its decay products. The size of the full off-shell effects is also estimated by an explicit comparison to the full Narrow-Width-Approximation (NWA) and to the NWA with LO decays.

\section{Results}
We consider the processes   $pp\rightarrow e^+ \nu_e \mu^-\overline{\nu}_{\mu} e^+ \nu_e b \overline{b} +X$ and $pp\rightarrow e^- \overline{\nu}_e \mu^+ \nu_{\mu} e^- \overline{\nu}_e b \overline{b} +X$ at NLO in QCD at the LHC with $\sqrt{s}=13$ TeV and provide predictions for the full off-shell, the full NWA and the NWA with LO decays using the Monte Carlo \texttt{HELAC-NLO} \cite{helac1}-\cite{helac2} software package. For ease of notation we will use respectively $t\overline{t}W^+$ and $t\overline{t}W^-$ to label the processes. The final state is required to consist of three charged leptons and exactly two b-jets. Jets are reconstructed out of all final state partons using the IR-safe anti-$k_T$ jet algorithm \cite{antikt} with the separation parameter $R=0.4$. The following cuts were imposed on all events :\\

\begin{tabbing}
\hspace{4cm}\=\hspace{4cm}\=\kill
\>$p_T(\ell)>25~\rm{GeV}$  \>  $p_T(j_b)>25~\rm{GeV}$\\
\>$|y(\ell)|<2.5$ \> $|y(j_b)|<2.5$\\
\>$\Delta R (\ell\ell) >0.4$ \>  $\Delta R (\ell j_b) >0.4$\nonumber
\end{tabbing}

\noindent where $\ell$ denotes charged leptons $\ell=e^+,\mu^-$. 
We use the 5-flavour scheme with the PDF-set \texttt{NNPDF3.0} \cite{nnpdf30} with $\alpha_S=0.118$. For the renormalisation and the factorisation scales we use a fixed scale choice $\mu_0 = m_t + m_W/2$
and a dynamic scale choice $\mu_0 = H_T/3$ with $H_T=p_T(\ell_1) +p_T(\ell_2)+p_T(\ell_3)+p_T(j_{b_1})+p_T(j_{b_2})+p_T^{miss}$. 

\begin{figure}
\begin{floatrow}
\ffigbox{%
  \includegraphics[scale=0.2]{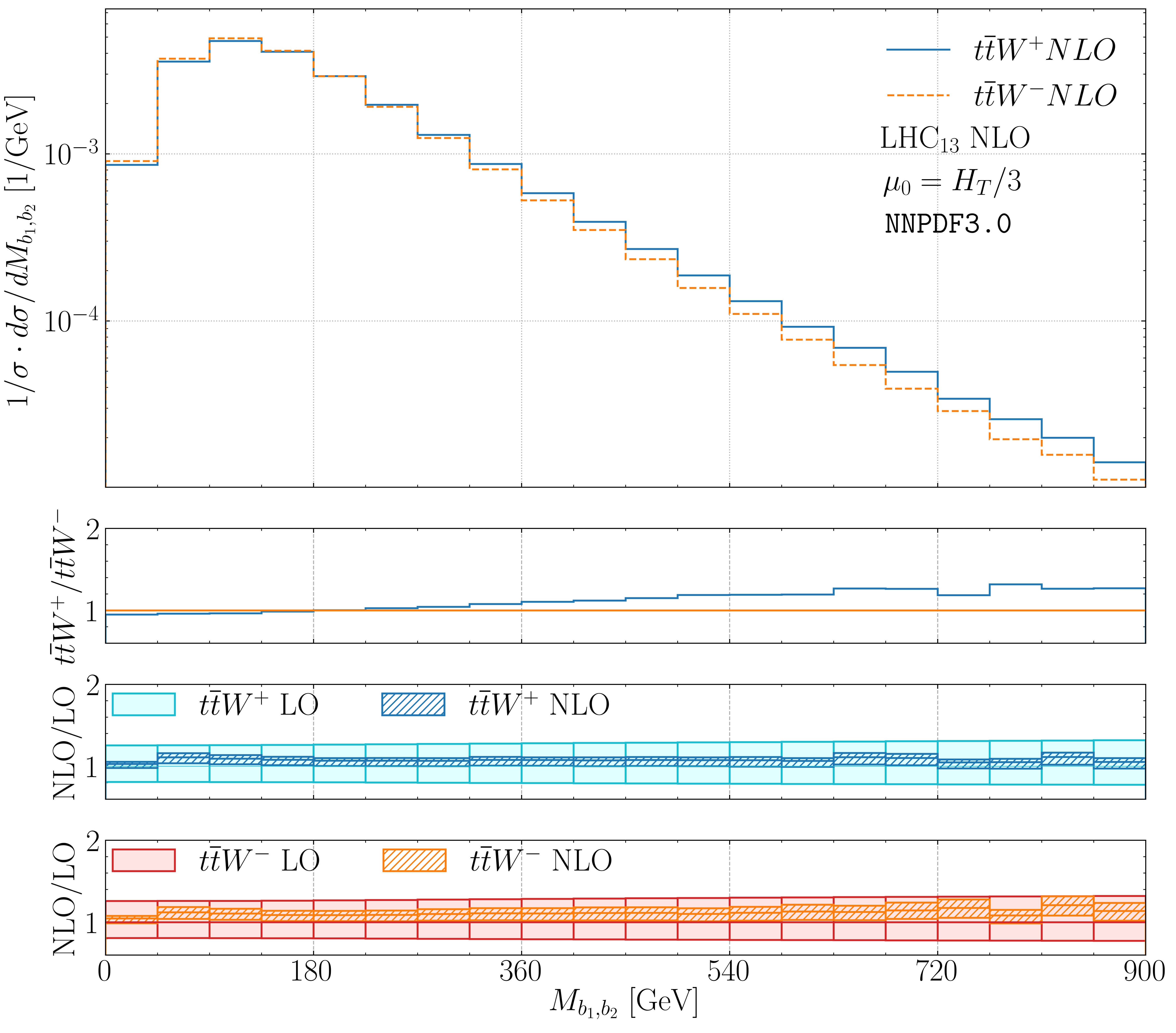}
}{%
  \caption{Differential distribution of the invariant mass of the b-jet system for the dynamical scale. }%
 \label{fig1}
}

\capbtabbox{%
\begin{tabular}{ llll } \hline
  & $\sigma^{\textcolor{blue}{t\overline{t}W^{+}}}_{\rm{NLO}}$ [ab] & $\sigma^{\textcolor{red}{t\overline{t}W^{-}}}_{\rm{NLO}}$ [ab] \\ \hline
 \textbf{Full off-shell}  & $124.4^{+3\%}_{-6\%}$ & $68.6^{+5\%}_{-7\%}$\\ 
 \textbf{NWA}   & $124.2^{+3\%}_{-6\%}$ & $68.7^{+5\%}_{-7\%}$\\
 \textbf{NWA$_{LOdec}$}     &$130.7^{+10\%}_{-10\%}$ & $72.0^{+11\%}_{-11\%}$\\\hline
 \\
 \\
 \\
\end{tabular}
}{%
  \caption{Integrated fiducial cross sections at NLO in QCD for various modelling approaches for the dynamic scale. }%
  \label{table1}
}
\end{floatrow}
\end{figure}

\noindent We begin the discussion by motivating the cross section ratio as an interesting observable to consider for $t\overline{t}W^{\pm}$. The two processes are similar at NLO when comparing several relevant quantities. First we note that the $\mathcal{K}$-factor for both is at the level of $\mathcal{K}\sim 1.1$. In Table \ref{table1} we show the fiducial integrated cross section at NLO for $t\overline{t}W^+$ and $t\overline{t}W^-$ for various modelling approaches. Yet another similarity can be observed from this table, namely that the size of the theoretical scale uncertainty is also similar between the two processes. For the full off-shell and the full NWA it is around $7\%$, whereas for the NWA with LO decays it increases to $10\%-11\%$.  The PDF uncertainty at NLO is also similar, at a level of $2\%$. Both processes have similar small off-shell effects, of the order of $0.1\%-0.2\%$. \\
We examine the similarities between $t\overline{t}W^+$ and $t\overline{t}W^-$ at the differential level as well, by taking a look at the invariant mass between the two b-jets $M_{b_1,b_2}$ in Figure \ref{fig1}. To this end, we show the normalised differential distribution for the central scale $\mu_0=H_T/3$ in the uppermost panel. The second panel shows the ratio  between these central lines whereas the third and fourth panels give the differential $\mathcal{K}$-factors for $t\overline{t}W^+$ and $t\overline{t}W^-$ respectively, as well as the corresponding theoretical scale error band. While the central lines are almost identical at the bulk of the distribution, they show a similar trend towards the tails as well. The two lowermost panels also show a similar mild change of the differential $\mathcal{K}$-factor between $t\overline{t}W^+$ and $t\overline{t}W^-$. \\
Thus, motivated by a thorough investigation of these similarities we can treat the two processes as correlated with regards to the scale choice. Based on this finding, we expect the integrated cross section ratio $\mathcal{R}\equiv \sigma^{NLO}_{t\overline{t}W^+}/\sigma^{NLO}_{t\overline{t}W^-}$ at NLO in QCD to exhibit enhanced perturbative stability and subsequently aim to increase the precision of NLO QCD predictions. Computing this observable for the full off-shell at NLO QCD yields:

\begin{eqnarray}
\mathcal{R} = 1.81 \pm 0.03~(\rm{scale}) \pm 0.03 (\rm{PDF})  
\end{eqnarray}

\noindent The theoretical scale and PDF uncertainties amount to $2\%$ each, which makes for a very precise prediction of an observable at NLO QCD. This result is proven to be robust with respect to different modelling approaches, consecutive increases of the $p_T(b)$ cut up to $40~\rm{GeV}$ and the choice of scale (fixed or dynamic). It makes for an excellent choice to be considered in comparisons with experimental data, with the goal of finding new physics.

\begin{figure}[htb!!!]
    \centering

        \includegraphics[width=0.5\textwidth]{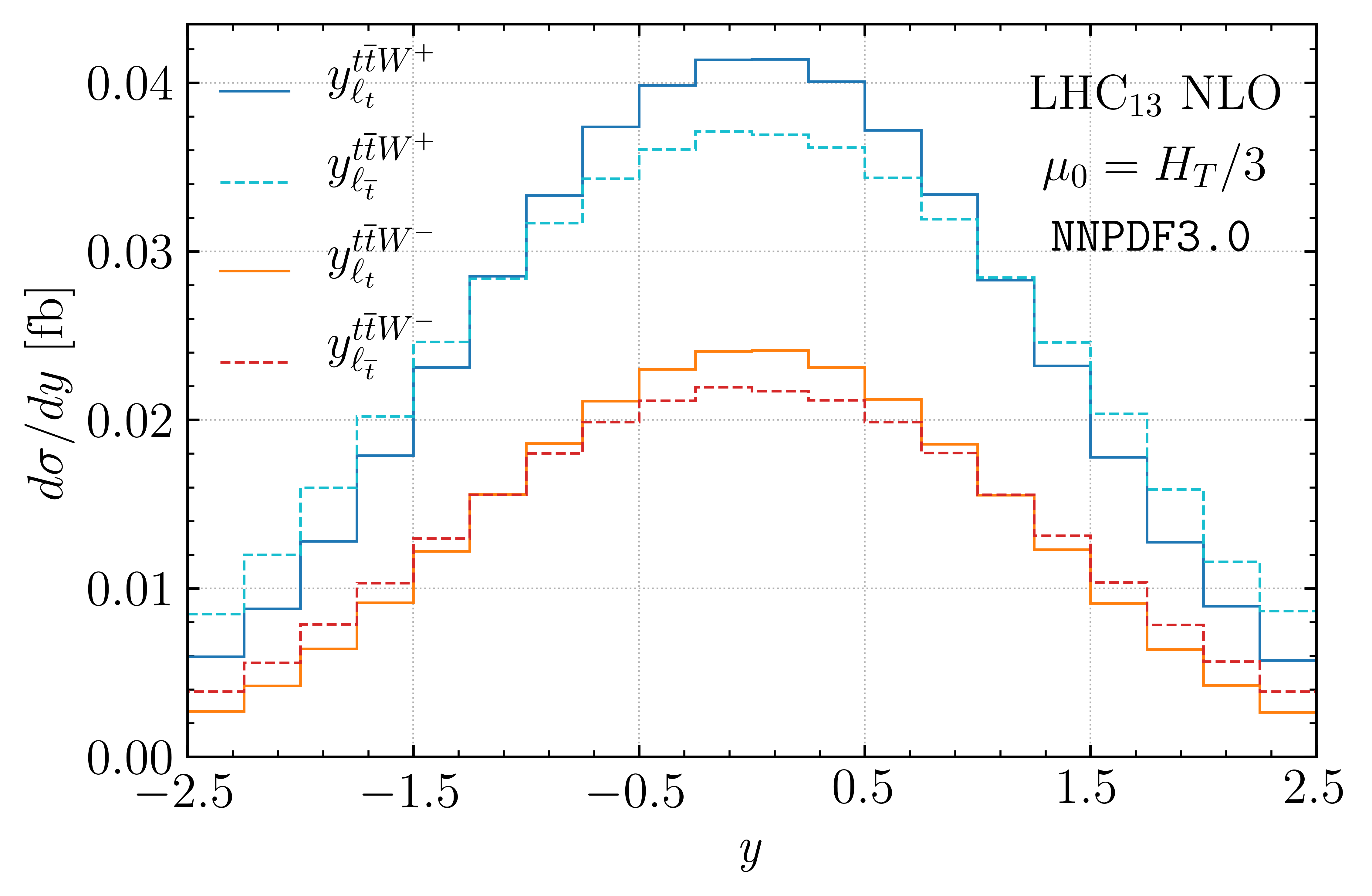} 
        \includegraphics[width=0.4\textwidth]{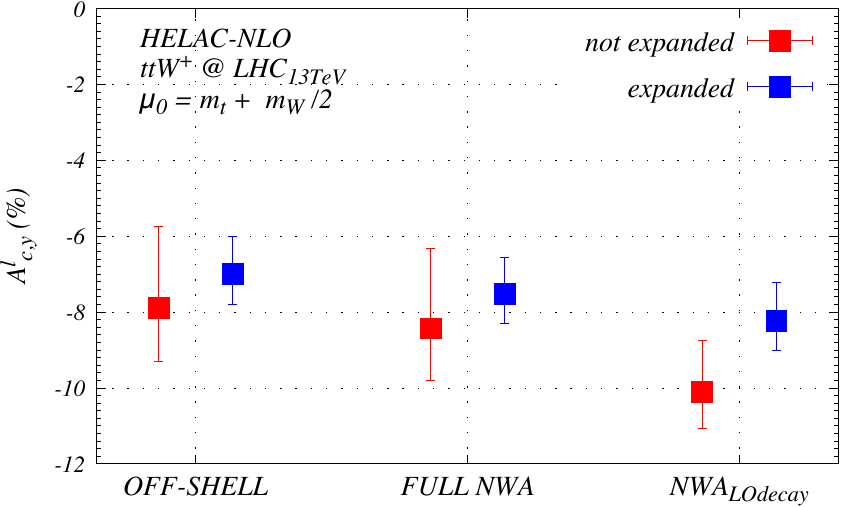} 
        \caption{Rapidity distributions of leptons from the top and anti-top for $t\overline{t}W^{+}$ and $t\overline{t}W^{-}$ (left) and expanded vs unexpanded charge asymmetries for the leptons from the top decays for $t\overline{t}W^{+}$ for various modelling approaches. }
        \label{fig2}
\end{figure} 

\noindent Another interesting set of observables for $t\overline{t}W^{\pm}$ are the charge asymmetries of the top and its decay products. As pointed out initially by \cite{mangano}, the top charge asymmetry for $t\overline{t}W^{\pm}$ is significantly larger than for $t\overline{t}$ because of the absence of the symmetric initial state $gg$ as well as the presence of the additional $W$-gauge boson which acts as a polariser to the tops. It can be nicely visualised by showing the rapidity distributions of the decay products from the top and anti-top on the same plot, as shown in Fig.\ref{fig2}. The charge asymmetry for the leptons originating from the top decay can be computed using:

\begin{eqnarray}
A_{c,y} = \frac{\sigma\left( \Delta|y| > 0 \right)-\sigma\left( \Delta|y| < 0 \right)}{\sigma\left( \Delta|y| > 0 \right)+\sigma\left(\Delta|y| < 0 \right)}~~~~~~~~\Delta|y| \equiv |y_{\ell_{t}}|-|y_{\ell_{\overline{t}}}|\nonumber 
\end{eqnarray}

\noindent We choose to give the final results in terms of expanded asymmetries, where we expand in  $\alpha_S$ and only take the contributions up to the first order. This removes the additional higher order contributions generated by the ratio, which can be affected by the unknown NNLO corrections. The expanded asymmetries with the full off-shell effects at NLO in QCD for the tops and the leptons from their decay are given as:

\begin{eqnarray}
A^{t}_{c,\eta,exp}~~[\%] =3.70^{+12\%}_{-11\%}~~~A^{t}_{c,y,exp}~~[\%]=2.62^{+15\%}_{-13\%}~~~A^{\ell}_{c,y,exp}~~[\%]=-7.00^{+14\%}_{-11\%}
\end{eqnarray}

The impact of the modelling and the comparison between expanded and unexpanded is shown in Fig.\ref{fig2} for the leptons from the tops. There is a notable shift upwards for the central value of the expanded charge asymmetries with respect to the unexpanded asymmetries as well as a reduction in the size of the theoretical scale error. Regarding the modelling, there is generally good agreement between the full off-shell and the full NWA, whereas the NWA with LO decays can show differences of up to $2\sigma$ for certain observables not shown here. 

\section{Conclusions}

We employ the state of the art full off-shell results in NLO QCD for $t\overline{t}W^{\pm}$ to study the cross section ratio and the charge asymmetries of the tops and their decay products. The contributions from the off-shell effects are assessed continuously through a comparison to the full NWA.
We also show results for the NWA with LO decays to get an idea of the impact of the NLO QCD corrections to the top decays. Based on the similarity between the processes $t\overline{t}W^{+}$ and $t\overline{t}W^{-}$
as far as NLO QCD corrections are concerned, we motivated the use of the integrated cross section ratio $\mathcal{R}\equiv \sigma^{NLO}_{t\overline{t}W^+}/\sigma^{NLO}_{t\overline{t}W^-}$. Its theoretical scale uncertainty at NLO is reduced to $2\%$, which is similar to the PDF uncertainty. This level of precision and its stability with respect to different modelling approaches, scale choice and minimum $p_{T}(b)$ cut suggest that $\mathcal{R}$ is a good choice to test the precision of the SM and the presence of new physics. \\
On the other side, the charge asymmetries for $t\overline{t}W^{\pm}$ can be altered in the presence of new physics as well. We revisit these and compute the expanded and unexpanded charge asymmetries with the full off-shell effects and compare to the other modelling approaches.

\section*{~~~~~~~~~~~~~~~~ACKNOWLEDGEMENTS}
This work was supported by the Deutsche Forschungsgemeinschaft (DFG) under grant 396021762 - TRR 257: P3H - Particle Physics Phenomenology after the Higgs Discovery and under grant 400140256 - GRK 2497: The physics of the heaviest particles at the Large Hadron Collider.

\end{document}